%% file: Yang.tex
\documentclass[9pt,conference]{IEEEtran}
\IEEEoverridecommandlockouts
\pdfoutput=1
\usepackage{cite}
\usepackage{amsmath,amssymb,amsfonts}
\usepackage{algorithmic}
\usepackage{graphicx}
\usepackage{textcomp}
\usepackage{xcolor}
\def\BibTeX{
{\rm B\kern-.05em{\sc i\kern-.025em b}\kern-.08em
    T\kern-.1667em\lower.7ex\hbox{E}\kern-.125emX}}

\usepackage{geometry}
\usepackage{threeparttable}
\usepackage{caption}
\usepackage{url}
\usepackage{cases}
\usepackage{bbm}
\usepackage{bm}

\usepackage{epsfig,epstopdf}

\usepackage{algorithm}
\usepackage{algorithmic}
\usepackage{booktabs}
\usepackage{multirow}
\usepackage{mathdots}
\usepackage{array}
\usepackage{stmaryrd}

\usepackage{graphicx}
\usepackage{subfigure}
\usepackage{enumitem}

\begin{document}

\title{Flexible Android Malware Detection Model based on Generative Adversarial Networks with Code Tensor
}

 \author{\IEEEauthorblockN{ Zhao Yang\IEEEauthorrefmark{2}\IEEEauthorrefmark{1}\thanks{\IEEEauthorrefmark{2} is the corresponding author. Email: lingxi.yz@alibaba-inc.com}}
 \thanks{\IEEEauthorrefmark{1} contribute equally to this
paper}
\IEEEauthorblockA{\textit{Alibaba Group} \\
Shenzhen, China \\
Email: lingxi.yz@alibaba-inc.com
}
\and
\IEEEauthorblockN{Fengyang Deng\IEEEauthorrefmark{1}}
\IEEEauthorblockA{
\textit{Huazhong University of Science and Technology} \\
Wuhan, China \\
Email: fengyang\_deng@hust.edu.cn}
\and
\IEEEauthorblockN{Linxi Han}
\IEEEauthorblockA{\textit{Xi’an International Studies University} \\
Xi’an, Shaanxi\\
Email: hanlinxi@sina.cn}
}

\maketitle

\begin{abstract}
   The behavior of malware threats is gradually increasing, heightened the need for malware detection. However, existing malware detection methods only target at the existing malicious samples, the detection of fresh malicious code and variants of malicious code is limited. In this paper, we propose a novel scheme that detects malware and its variants efficiently. Based on the idea of the generative adversarial networks (GANs), we obtain the `true' sample distribution that satisfies the characteristics of the real malware, use them to deceive the discriminator, thus achieve the defense against malicious code attacks and improve malware detection. Firstly, a new Android malware APK to image texture feature extraction segmentation method is proposed, which is called segment self-growing texture segmentation algorithm. Secondly, tensor singular value decomposition (tSVD) based on the low-tubal rank transforms malicious features with different sizes into a fixed third-order tensor uniformly, which is entered into the neural network for training and learning. Finally, a flexible Android malware detection model based on GANs with code tensor (MTFD-GANs) is proposed. Experiments show that the proposed model can generally surpass the traditional malware detection model, with a maximum improvement efficiency of 41.6\%. At the same time, the newly generated samples of the GANs generator greatly enrich the sample diversity. And retraining malware detector can effectively improve the detection efficiency and robustness of traditional models.
\end{abstract}


\section{Introduction}
The mobile Internet and intelligent mobile devices have undergone rapid development in the past decade, however they bring us security risks because of malware at the same time. About 80\% of smartphone users are using the Android operating system in the world. According to the report by G Data \cite{GdataReport}, the total number of mobile malware increased by 40\% in 2018, 3.2 million new Android malware samples were detected by the end of the third quarter of 2018. The threat from the Android system has reached a new level.

Traditionally, malicious code analysis methods include both static analysis \cite{Christodorescu2005SemanticsawareMD} and dynamic analysis \cite{Yin2007PanoramaCS}. Currently, researchers extracted the features of malware, such as the call sequence of API functions, permissions requested, etc. and then analyzed them using machine learning methods. Ye et al. \cite{Ye2017ASO} extracted the API call sequences of malware, control flow graph, and other features, used information chain, word frequency statistics and other methods. Cui et al. \cite{Cui2019MaliciousCD} detected malicious code based on CNNs and multi-objective algorithm, they converted binary executable files of malware to a grayscale image, then combined the CNNs to detect malware.

However, static analysis is susceptible to code obfuscation techniques, and the dynamic analysis detection results may be affected by omitting key executable paths. Ye and Cui et al. analyzed the entire executable file of the malware, which tends to weaken the malicious  features of the code. Moreover, These methods are based on existing samples of malware and results in the detection of malware certain hysteresis.

Regarding the problems above, there are several challenges we have to attach importance to:

Firstly, how to enhance the malicious features. Considering a mix of irrelevant features can weaken themselves, separating a complete executable filecode into feature fragments with malicious behavior is necessary.

Secondly, how to unify the size of the feature after segmentation and minimize the behavior pattern loss when adjusting the feature size. According to the current feature extraction methods, an effective code feature decomposition or mapping need to be designed.

Thirdly, how to detect new malware and variations of existing malware. If new malware features can be generated according to the existing malware samples, it can not only enrich the malicious samples, but also improve the efficiency of active defense.

\par To address the above challenges, firstly, we map binary malicious code to image, secondly segment the image using the self-growing segmentation algorithm, thirdly use the tensor singular value decomposition to transform the different size segments into third-order tensors. Finally based on the GAN idea, we generate new malicious code samples and improve the detector performance. Based on the idea of active defense and the purpose of upgrading the original malicious code detector model, Goodfellow et al. \cite{Goodfellow2014GenerativeAN} proposed GAN (Generative Adversarial Network), which adopts the idea of adversarial generation and build a network composed of generator and discriminator, and the model is trained by adversarial learning. In the field of information security, GAN's development mainly focuses on obtaining adversarial samples \cite{Xiao2018GeneratingAE} and generating adversarial virus samples \cite{Hu2017GeneratingAM}. Based on these theories, we use the generative adversarial network to generate adversarial samples in the setting environment of a semi-white-box attack and black-box attack.
\par In this paper, we propose a novel malware detection method that utilizes the idea of GANs to generate `true' samples satisfying the distribution characteristics of malware data and repeatedly trains the malware detector model. It can effectively enrich the dataset of unknow malware samples, resist the active attack of malware, and improve the detection of efficiency of malware. 

\par The main contributions of this paper are as follows:

Firstly, a novel method of segmentation from self-growing malware APK to texture image features is proposed. We map the binary malicious code segments into images and analyze the malicious code segments based on the image texture. We design the texture cutting algorithm based on the Locality Sensitive Hashing (LSH) algorithm to extract significant feature texture segments from malicious code texture segments and enhance the texture features of malware. 

Secondly, the Singular Value Decomposition (SVD) based on Low-tubal rank is used to strengthen the characteristics of malicious code. The images of different sizes are unified into a fixed-size third-order tensor as the input of the neural network model. 

Thirdly, a flexible malware detection framework MTFD-GANs based on the anti-generation network is proposed. New malicious code features are generated in the training model, they enrich the diversity of samples and enhance the robustness of the model. We extracted 2000 data with obvious feature types from the Drebin dataset for testing. The experimental results show that the proposed model outperforms the traditional malware detection model, with the maximum improvement efficiency 41.6\%. 

\par The main structure of this paper is as follows. Section 2 presents the background. Section 3 details the preprocessing for binary code and the structure of MTFD-GANs. Section 4 introduces the training of MTFD-GANs. Section 5 verifies the validity of our proposed model through experiments. Finally, Section 6 concludes this paper.

\section{Background}
This section first introduces the Locality Sensitive Hashing algorithm used for significant feature segment extraction. Then detail the principle of tensor singular value decomposition. Finally, the Black-Bone prediction model is described.

\subsection{Locality Sensitive Hashing}
Locality Sensitive Hashing (LSH) \cite{Gionis1999SimilaritySI} is based on the idea that, multiple hash functions are used to project large-scale high-dimensional data points, so that the closer the points are, the more likely they remain close together, and vice versa. Let $x$ and $y$ be two different high-dimensional feature vectors. In LSH index algorithm, the probability of remaining close is usually related to the similarity, that is:
\begin{equation}\label{eq:lsh-probality}
    Pr_{hj \in H}[h_j(x)=h_j(y)]=sim(x,y)
\end{equation}

Where $H$ is the hash function cluster, $h_j$ is the hash function randomly selected in the hash function cluster, and $sim()$ is the similarity function.

Obviously, LSH algorithm depends on locally sensitive hash function family. Let $H$ be a hash function family mapped by $R^d$ to set $U$. For any two points $p$ and $q$, a hash function $H$ is randomly selected from the hash function family $H$. If the following two properties are satisfied, the function family $H={h:R^d \to U}$ is called $(r_1, r_2, p_1, p_2)$ locally sensitive:

\begin{itemize}
\item if $D(p, q) \le r_1$, then $Pr_H[h(p)=h(q)] \ge p_1$,
\item if $D(p, q) \le r_2$, then $Pr_H[h(p)=h(q)] \ge p_2$.
\end{itemize}

Where $r_1 < r_2$ and $p_1 > p_2$. The function of LSH function family is used for hashing, which can ensure that the collision probability of the close points is greater than that of the far points.

\subsection{Low-tubal-rank Tensor}
We use lowercase letters to represent scalar variables, e.g., $x$, $y$, and bold lowercase letters to indicate vectors, e.g., $\bm{x}$, $\bm{y}$.  The matrix is represented by bold uppercase letters, e.g., $\bm{X}$, $\bm{Y}$, and higher-order tensor is represented by calligraphic letters, e.g., $\mathcal{X}$, $\mathcal{Y}$. The transposition of high-order tensor is indicated by the superscript $^\dagger$, e.g., $\mathcal{X}^{\dag}$, $\mathcal{Y}^{\dag}$, which first transposes the elements of all the previous slice matrices and then reverses the order of the slices, from the $2$-th slice to the $I_3$-th slice. In order to calculate the clarity of the description, we define the tensor $\widetilde{\mathcal{T}}$ mapped by the frequency domain space to represent the original tensor $\mathcal{T}$ to perform Fourier transform along the third dimension.

\textbf{Tubes/fibers and slices of a tensor}: The higher-order analogue of a matrix's column is called tube, which is defined by a one-dimensional fixed direction. $\mathcal{T}(:,j,k)$, $\mathcal{T}(i,:,k)$ and $\mathcal{T}(x,j,:)$ are used to represent mode-1, mode-2, and mode-3 tubes, respectively, which are vectors. While a slice is defined by a two-dimensional matrix, $\mathcal{T}(:,:,k)$, $\mathcal{T}(:,j,:)$ and $\mathcal{T}(i,:,:)$ represent the front, lateral, horizontal slices, respectively. In addition, if all the front slice matrices of the tensor are diagonal, then call it $\mathnormal{f}$-diagonal tensor.

\textbf{t-product} \cite{Kilmer2013Third, kilmer2011factorizationGoogle}: Let $\mathcal{A}$ is $I_1\times I_2\times I^{'}$, $\mathcal{B}$ is $I_2\times I_3\times I^{'}$, the t-product of $\mathcal{A}$ and $\mathcal{B}$ can be expressed as
	\begin{equation}\label{eq:t-procuct}
		\mathcal{A}\ast \mathcal{B} = \text{fold} \big(\text{circ}(\mathcal{A})\cdot \text{MatVec}(\mathcal{B})\big),
	\end{equation}
	where circ($\mathcal{A}$) is the circular matrix of tensor $\mathcal{A}$, and MatVec($\mathcal{B}$) is the block $I_2I^{'}\times I_3$ matrix that is obtained by tensor $\mathcal{B}$. In this paper, the product of two tensors, also called the tensor circular convolution operation.

\textbf{Third-order tensor block diagonal and circulant matrix} \cite{Kilmer2013Third, kilmer2011factorizationGoogle} : For a third-order tensor $\mathcal{A} \in \mathbb{R}^{I_1 \times I_2 \times I_3}$, we denote the block $\bm{A}^p\in \mathbb{R}^{I_1 \times I_2}$ as the matrix obtained by holding the third index of $\mathcal{A}$ fixed at $p$, $p\in [I_3]$ in the Fourier domain. The block diagonal form of third-order tensor $\mathcal{A}$ can be expressed as
	\begin{equation}\label{eq:block diagonal}
		blkdiag(\widetilde{\mathcal{A}}) \triangleq 
		\left[\begin{array}{ccccc}
		\widetilde{\bm{A}}^1 & & & \\
 		& \widetilde{\bm{A}}^2& &  \\
		& &\ddots & \\
		& & & \widetilde{\bm{A}}^{I_3}
		\end{array}
		\right]\in \mathbb{C}^{I_1 I_3 \times I_2 I_3}
	\end{equation}
	where $\mathbb{C}$ denotes the set of complex numbers.
	 We use the MatVec$(\cdot)$ function to expand the front slices of the tensor
  \begin{equation}\label{eq:MatVec}
  \text{MatVec}(\mathcal{A}) = \left[
  \begin{array}{cccc}
  \bm{A}^1 \\
  \bm{A}^2 \\
  \vdots \\
  \bm{A}^{I_3}
  \end{array}
  \right]\in \mathbb{R}^{I_1I_3\times I_2}.
 \end{equation}
 The operation takes MatVec($\mathcal{A}$) back to the form of the original tensor by 
\begin{equation}\label{eq:fold}
	\text{fold}(\text{MatVec}(\mathcal{A})) = \mathcal{A},
\end{equation}
then the circulant matrix of third-order tensor $\mathcal{A}$ is
  \begin{equation}\label{eq:bcirc}
  \begin{split}
  \text{circ}(\mathcal{A}) = \left[
  \begin{array}{cccc}
  \bm{A}^{1} & \bm{A}^{I_3} & \cdots & \bm{A}^{2} \\
  \bm{A}^{2} & \bm{A}^{1} & \cdots & \bm{A}^{3} \\
  \vdots   & \vdots   & \ddots & \vdots \\
  \bm{A}^{I_3} & \bm{A}^{I_3 - 1} & \cdots & \bm{A}^{1}\\
  \end{array}
  \right]\in \mathbb{R}^{I_1I_3\times I_2I_3}.
  \end{split}
  \end{equation}

\subsection{Black-Bone Detector}
\par We use the Black-Bone detector to detect samples generated by GANs, and iteratively update the generator and discriminator based on the detection results of the Black-Bone detector. The malware detection models used in Black-Bone include Support Vector Machine, Logistic Regression, Decision Tree, Random Forest, Multi-Layer Perceptron, Attention, AdaBoost, Gradient Boosting Decision Tree, Naive Bayes, etc. 

Support Vector Machine (SVM) \cite{Cortes1995SupportvectorN} is a supervised learning model with associated learning algorithms, and is usually used for classification analysis. Logistic Regression (LR) \cite{Pampel2000LogisticRA} models a binary dependent variable using a logistic function, such as the probability of a certain class. Decision Tree \cite{Safavian1991ASO} is a prediction model, represents a mapping relationship between object attributes and object values. Random Forest \cite{Breiman2001RandomF} is a classifier that constructing multiple decision trees to train and predict, corrects for overfitting caused by a single decision tree. Multi-Layer Perception \cite{Hastie2005TheEO} is a feedforward artificial neural network model, maps a set of input vectors to a set of output vectors. Attention \cite{Bahdanau2015NeuralMT} draws on the mechanism of signal processing of the human brain. After scanning the whole image, it determines the target area worthy of attention and then identifies this target area in a more detailed way to eliminate useless information and extract more meaningful details. AdaBoost \cite{Freund1995ADG} trains different classifiers (weak classifiers) for the same training set, and then assemble these weak classifiers to form a stronger final classifier (strong classifier). Gradient Boosting Decision Tree (GBDT) \cite{Friedman1991MultivariateAR} uses the idea of iteration to reduce the residuals generated during training, consists of multiple decision trees, and the conclusions of all decision trees are combined to make the final answer. Naive Bayes (NB) \cite{Maron1961AutomaticIA} classifier is based on Bayes' theorem and assumes that the feature conditions are independent of each other.

\section{MTFD-GANs}
This section first introduces the overall framework for malware  detection based on generative adversarial networks (GANs). Secondly, it is introduced from the processing of binary code to the discriminator and generator core in the GANs network.

\subsection{Overview}
\begin{figure*}[htb]\small
	\centering
	\includegraphics[width=0.8\textwidth]{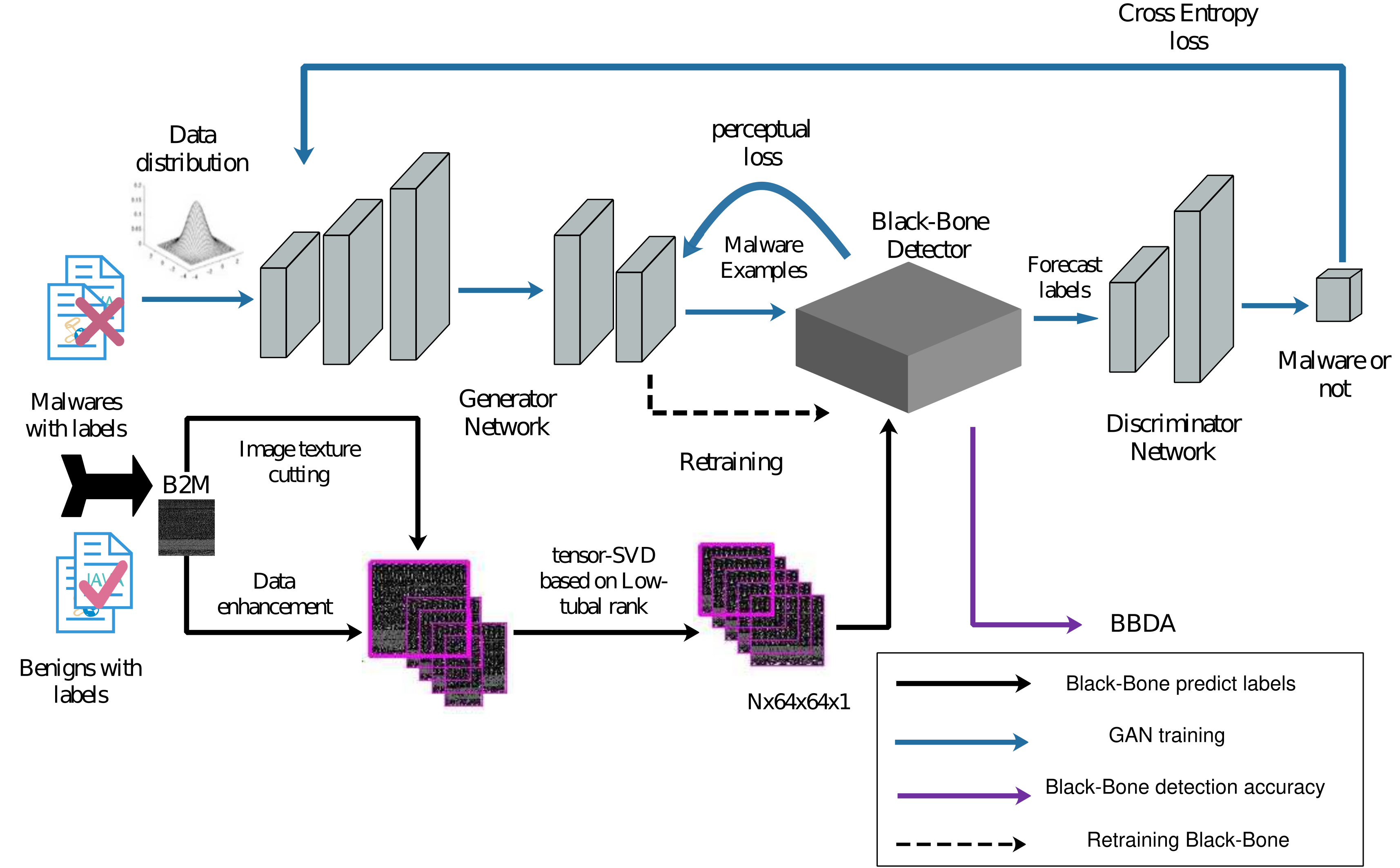}
	\caption{Malicious code detection model based on GANs. The Black-Bone detection accuracy (BBDA) means the detection rate of malware.}
	\label{fig:framwork}
	\vspace{-8pt}
\end{figure*}

The general Android malware detection idea is based on the extraction of operation code (opcode) sequences or Application Programming Interface (API) features and establishes a classified neural network for training. However, in order to protect the relationship between data as much as possible, we propose a code-to-image lossless coding method and based on image texture feature extraction and t-SVD, to achieve uniform image size. Fig. \ref{fig:framwork} shows the established flexible malware detection framework for code tensor. First, an executable binary Android software is losslessly encoded into a grayscale image using the B2M (Binary mapping to image) algorithm \cite{nataraj2011malware}. Second, the original large image is cut using a segment self-growing image texture segmentation algorithm. On the one hand, in order to extract significant feature texture segments based on the LSH algorithm \cite{abdulhayoglu2018use}, on the other hand, it plays a role in data enhancement and efficiency. Finally, based on the singular value decomposition (SVD) method of tensor-based Low-tubal rank, the texture image and the third-order tensor compressed to the uniform size are input into the neural network model.

The flexibility of the entire framework is easy to migrate and add to the existing malicious code detection framework. It enhances the detection efficiency and robustness of the model and plays an important role in resisting malicious code attacks. The framework of the entire model is divided into three main parts: the discriminative network, the generator network, and the Black-Bone prediction. The generative adversarial network (GAN) is a supervised learning process based on the addition of an auxiliary classifier model.

\subsection{Preprocessing for Binary Code}
This section introduces the key technologies involved in the paper, including image texture cutting technology and eigenvalue decomposition of tensor, to achieve feature value extraction and compression for code tensor.

\textbf{Binary to Image}: There is an executable binary file in each Android software. We extract the binary dex file from Android APK, convert it into a gray image using B2M(Binary mapping to Image) algorithm. It reads an 8-bit unsigned form (between 0 and 255) each time, and finally generates a two-dimensional matrix after lossless coding. Each pixel value is between 0 and 255. 

We fix the width of the two-dimensional matrix to 256 pixels, and the height of the matrix varies with the size of the binary file.

\textbf{Image Texture Cutting}: After getting the gray images, we analyze them according to the texture features of the image. But important information would be lost if the original image is compressed directly. Therefore, we cut the image with a similar texture before processing, making the texture features of malicious code segments more obvious. The difficulty of image texture segmentation is mainly reflected in the roughness, smoothness, directionality, and regularity of image texture.
\par Common texture segmentation algorithms are mainly based on wavelet transform \cite{Wang2003SupervisedTS, Charalampidis2002WaveletbasedRI}, Gabor filtering \cite{Pichler1998AnUT}, and self-growing. The image texture segmentation algorithm we use is based on self-increasing similarity attribute distance calculation. 
\par We use $P(R_i)$ for the formal description of all the elements in set $R_i$, and $\phi$ for the empty set.
\begin{itemize}
\item [(1)] $R_i$ is a connected domain. After the segmentation of image texture, each segment must be in a corresponding domain, and the points in the subdomain must be connected. Segments with the same texture feature are divided into the same texture block, and segments with different texture features are divided into different subdomains.
\item [(2)] $R_i \cap R_j = \phi, i, j = 1, 2, \dots, N$, represents that the corresponding domains are disjoint, indicating that the malicious code textures do not contain the same features.
\item [(3)] $P(R_i)$ = True, $i = 1, 2,\dots, N$, represents that all subsegments in the same domain have similar texture features, and are merged into the same texture segment.
\item [(4)] For neighborhoods $R_i$ and $R_j$, $P(R_i \cup R_j)$ = False, indicating that there are distinct differences among adjacent texture subsegments and they do not belong to the unified texture segment.
\end{itemize}

\par The specific steps of the image texture cutting algorithm are shown in Alg. \ref{algorithm:ImageCut}
\begin{algorithm}[htb]
\caption{Image Texture Cutting}\label{algorithm:ImageCut}
\begin{algorithmic}[1]
\REQUIRE $P$, image of an Android APK; $P_i$, cell segment of image $P$ (two lines consist one cell segment); $N$, texture segment number of image $P$; $x_0$, texture segment currently processed; $x_1$, next neighborhood texture segment of $x_0$; $F_i$, state of texture segment $i$, if False, means segment $i$ is undisposed.
\STATE Initialize the flag of each texture segment $F_i$ = False
\FOR{$i$= 1 to $N$}
	\IF{$F_i$ is False}
		\STATE set $x_0 = P_i$
		\IF{not Degraded($x_0$) and SelfGrowing($x_1$)}
			\STATE Concatenate($x_0$, $x_1$)
		\ENDIF
		\STATE set $F_i$ = True
	\ENDIF
\ENDFOR
\ENSURE $P^{'}$, the set of cutting image.
\end{algorithmic} 
\end{algorithm}

\par when Degraded($x_0$) is True means texture segment $x_0$ satisfies the Degraded Criterion, while SelfGrowing($x_1$) is True means texture segment $x_1$ satisfies the Self-Growing Criterion. Where Concatenate($x_0$, $x_1$) represents domains of subsegment $x_0$ and subsegment $x_1$ are merged into the same domain.
\par Degraded Criterion: If the current texture segment is constant, the subdomain stops growing and ``disposed''  will be marked, and then delete the texture segment.
\par Self-Growing Criterion: If the feature distance between the current texture segment and the neighborhood texture segment is less than the given threshold (0.05 is setted in this paper), current texture segment and neighborhood texture segment will be merged; otherwise, the growth will be stopped and ``disposed'' will be marked, and the current texture segment is taken as new domain's starting segment.
\par We use the Entropy, Contrast, Homogeneity, and ASM (Angular Second Moment) of the texture segment's GLCM (Gray Level Co-occurrence Matrix) \cite{Haralick1973TexturalFF} as texture features of segments. And the Euclidean distance between two texture segments is used to calculate the feature distance between the texture segment and the neighborhood texture segment.
\par
\begin{figure}[htb]
\centering
	\subfigure[Implementation process]{
		\begin{minipage}[t]{0.46\linewidth}
		\centering
		\includegraphics[width=1.5in]{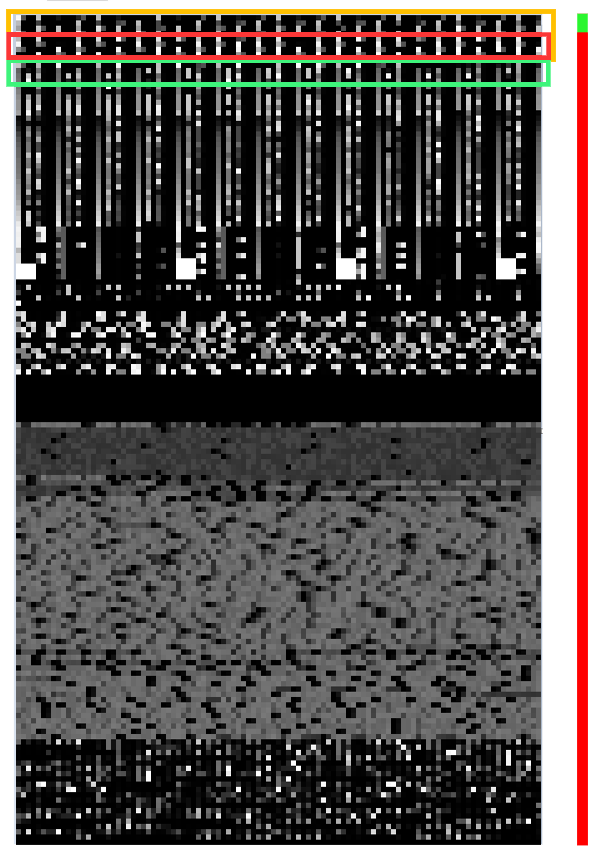}
		\label{fig:imgcut:a}
		\end{minipage}
	}
	\subfigure[The final segmentation result]{
		\begin{minipage}[t]{0.46\linewidth}
		\centering
		\includegraphics[width=1.5in]{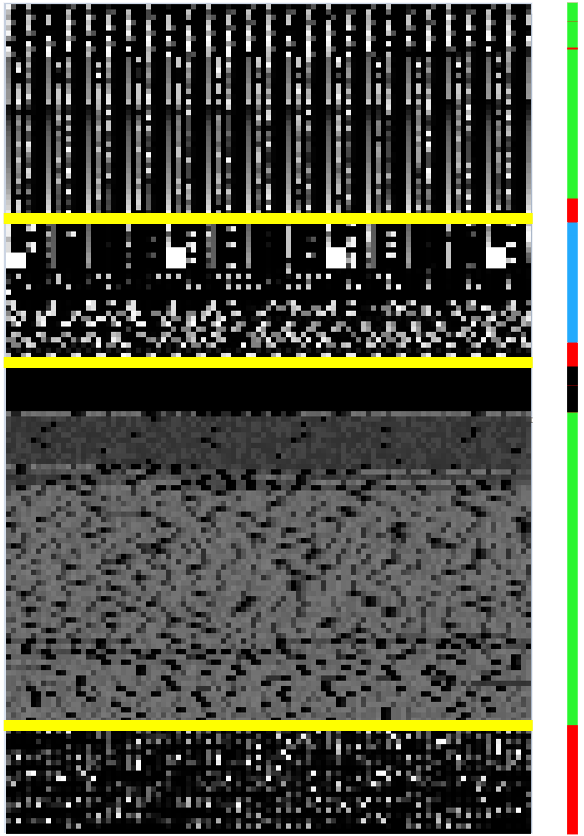}
		\label{fig:imgcut:b}
		\end{minipage}
	}
\centering
\caption{ Image texture cutting algorithm implementation process and final segmentation result.}
\end{figure}

 The image texture cutting algorithm execution process is shown in Fig.  \ref{fig:imgcut:a}. The final segmentation result of the texture image is shown in Fig.  \ref{fig:imgcut:b}.

\par For the cut images, the images whose height is less than 64 pixels are invalid.

\textbf{Significant feature segment extraction}: Some code segments are the part of public code segments of APK, while some code segments belong to some specific malicious behaviors, so the number of images with similar texture segments may be inconsistent after image cutting. To unify and balance the number of different types of texture images, we use LSH to extract conspicuous feature segments from the cut images. Different from the general hashing algorithm, LSH is based on location sensitivity, and the similarity between before and after hashing can be maintained to some extent.
\par We search and match image texture based on the LSH algorithm, look for a reliable texture block. A variety of hash functions are used to map image texture blocks to low-dimensional space, according to the distribution and features in the different hash table of image texture block, and low-dimensional spatial coding is used to represent high-dimensional data features. While in image searching and matching, Euclidean distance or Hamming distance is used.
\par We define a function cluster $G={g:S\to U}$, where $g(v)$ = $(h_1(v)$, $\dots$, $h_k(v))$, select K hash function $g_1, \dots, g_k$ from $G$ independently and randomly. For each point V in the dataset, store it in the bucket $g_i(v)$, where $i=1,2,\dots,l$. 
\par For an image texture feature vector $q$ to be searched, the given distance threshold $r$, we search the bucket $g_1(q),\dots,g_k(q)$, take all feature vectors $v_1,\dots, v_k$, as candidate approximate nearest distance. For any feature vector $v_j$, if $D(q, v_j)\le r$, then $v_j$ is returned, where $D$ is Euclidean distance between two element. The specific algorithm is shown in Alg. \ref{algorithm:LSH}.

\begin{algorithm}[htb]
\caption{LSH-based Image Search}\label{algorithm:LSH}
\begin{algorithmic}[1]
\REQUIRE texture image $match\_img$, texture image library $file$.
\STATE $\mathbf{function}$ $Lsh\_Search$($func$, $file\_list$, $hash$, \\ $hash\_id$, $character$, $match\_img$)
	\STATE $match\_lsh \gets HashMap(get\_vec(match\_img))$
	\FOR{$hash_i$ in $hash$}
		\IF{$match\_lsh$ in $hash_i$}
			\STATE $result\_imgs \gets file\_list_i$
		\ENDIF
	\ENDFOR
\STATE $\mathbf{end\ function}$
\FOR{$img_i$ in $file\_list(file)$}
	\STATE $p \gets get\_vec(img_i)$
	\STATE $character \gets p$
	\STATE $hash \gets HashMap(p)$
	\STATE $hash\_id \gets i$
\ENDFOR
\STATE $result\_imgs \gets Lsh\_Search$($func$, $file\_list$, $hash$, $hash\_id$, $character$, $match\_img$)
\ENSURE matched image $result\_imgs$
\end{algorithmic} 
\end{algorithm}

\textbf{Uniform feature size}: After image texture cutting and significant feature segment extraction, the size of the feature matrix would be inconsistent. In order to reduce the loss in the process of uniform feature size, the tSVD algorithm is used. The tSVD algorithm is carried out for the features whose size of the feature matrix is not less than 64$\times$64$\times n$, and the features are unified into 64$\times$64$\times$1 size after decomposition.

\textbf{t-SVD} \cite{Kilmer2013Third, kilmer2011factorizationGoogle}: The t-SVD of $\mathcal{T} \in \mathbb{R}^{I_1\times I_2\times I_3}$ is decomposed into $ \mathcal{T} = \mathcal{U} \ast \mathcal{S} \ast \mathcal{V}^{\dag}$, where the sizes of $\mathcal{U}$ and $\mathcal{V}$ are respectively  $I_1\times I_1\times I_3$ and $I_2\times I_1\times I_3$, $\mathcal{S}$ is a $\mathnormal{f}$-diagonal tensor of size $I_1\times I_2\times I_3$. The specific t-SVD algorithm is shown in Alg. \ref{algorithm:t-svd}

\begin{figure}[htb]\small
  \centering
  \includegraphics[width=0.48\textwidth]{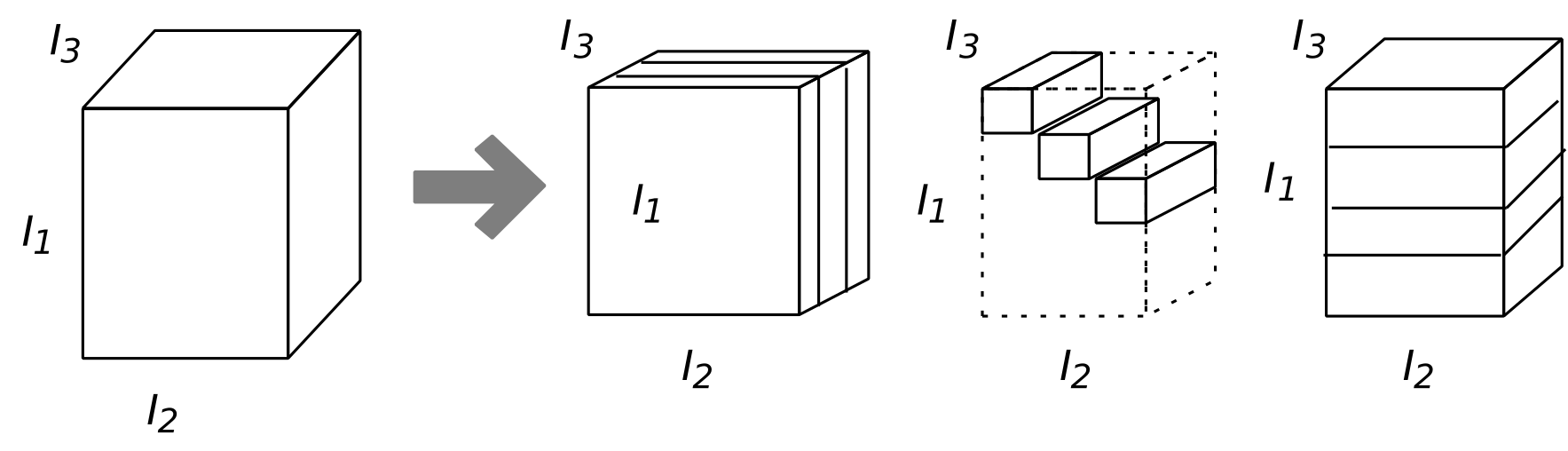}
  \caption{t-SVD of third-order tensor $\mathcal{T} \in \mathbb{R}^{I_1\times I_2\times I_3}$}\label{fig:tsvd}
\end{figure}
\begin{algorithm}[htb]
\caption{t-SVD \cite{Kilmer2013Third, kilmer2011factorizationGoogle}}\label{algorithm:t-svd}
\begin{algorithmic}[1]
\REQUIRE $\mathcal{T} \in \mathbb{R}^{I_1\times I_2\times I_3}$
\STATE $\widetilde{\mathcal{T}}$ $\leftarrow$ fft ($\mathcal{T}$,$[\;]$,3)
\FOR{i = 1 to $I_3$}
    \STATE $[\widehat{\bm{U}}, \widehat{\bm{S}}, \widehat{\bm{V}}]$ = SVD$(\widetilde{\mathcal{T}}(:,:,i))$
    \STATE $\widetilde{\mathcal{U}}(:,:,i)=\widehat{\bm{U}}$, $\widetilde{\mathcal{S}}(:,:,i)=\widehat{\bm{S}}$; $\widetilde{\mathcal{V}}(:,:,i)=\widehat{\bm{V}}$
\ENDFOR
\STATE $\mathcal{U}$ $\leftarrow$ ifft ($\widetilde{\mathcal{U}}$,$[\;]$,3); $\mathcal{S}$ $\leftarrow$ ifft ($\widetilde{\mathcal{S}}$,$[\;]$,3); $\mathcal{V}$ $\leftarrow$ ifft ($\widetilde{\mathcal{V}}$,$[\;]$,3)
\ENSURE $\mathcal{U}$, $\mathcal{S}$ and $\mathcal{V}$ $\hspace{2mm}$
\end{algorithmic} 
\end{algorithm}

We construct the tensor $\mathcal{T}$ by reading the feature images' pixel in the form of RGB, and take it as the input of t-SVD algorithm.

\textbf{Best rank-r approximate decomposition} \cite{Kilmer2013Third, kilmer2011factorizationGoogle}: Let the t-SVD of $\mathcal{A}\in \mathbb{R}^{I_1\times I_2\times I_3}$ be given by $\mathcal{A}=\mathcal{U}*\mathcal{S}*\mathcal{V}^{\dag}$ and for $k$ $<$ min$(I_1, I_2)$ define
	\begin{equation}\label{Tensor S-rank equation}
	\mathcal{A}_k=\sum _{i=1}^{k}\mathcal{U}(:,i,:)\ast \mathcal{S}(i,i,:)\ast \mathcal{V}^{\dag}(:,i,:).
	\end{equation}
	Then $\mathcal{A}_k$ = $\text{arg}\underset{\mathcal{A}\in \mathbb{M}}{\text{min}}$$||\mathcal{A} - \widehat{\mathcal{A}}||_{\digamma}$, where $\widehat{\mathcal{A}}$ is is the reconstructed tensor calculated after SVD decomposition and $\mathbb{M}$ = $\{\mathcal{C} = \mathcal{X}\ast \mathcal{Y}|\mathcal{X}\in \mathbb{R}^{I_1\times k\times I_3}, \mathcal{Y}\in \mathbb{R}^{k\times I_2\times I_3}\}$.

The output $\mathcal{U}$, $\mathcal{S}$ and $\mathcal{V}$ of t-SVD, we compress them by taking the first 64 feature vectors. After the above decomposition and dimension reduction, for each APK, a $64\times64\times n$ tensor can be obtained, and each slice $64\times64\times1$ of the tensor is taken out as the feature matrix of each APK. Finally, these feature matrixes are sent to the deep neural network for training or testing.

\subsection{Discriminator}
The training data of the discriminator consist of malware examples from the generator, and benign software is collected by malware authors. The ground-truth labels of the training data are not used to train the substitute detector. The goal of the substitute detector is to fit the Black-Bone detector. Therefore, the Black-Bone detector will first predict whether the label of the training data is malware or benign software and is used directly for the tag input of the discriminator.

The substitute detector is a multi-layer convolutional block with weights $\theta_{d}$ and a variety of strides which takes a program code tensor $\mathcal{X}\in \mathbb{R}^{64\times 64\times 1}$ as input in Tab \ref{table:Discriminator}. Each completed convolution block consists of a convolution layer \cite{lecun1998gradient-based}, instance normalization layer \cite{ioffe2015batch}, and LeakyReLU activation \cite{xu2015empirical}. The advantages of the Batch Normalization (BN) layer can also speed up the training, improve the generalization ability of the model, and disturb the training sequence of the sample. The BN layer is actually a normalized layer, which replaces the Local Response Normalized (LRN) \cite{robinson2007explaining}. If you put it after the activation function, at the beginning of the training, the interface is still shaking sharply and not stable enough, so after the basic mapping of the neuron unit, you can get stable results. The discriminator finally uses a two-layer fully connected network, the purpose of which is to play a smooth data feature mapping, and use the sigmoid activation function to predict the classification result to match the probability distribution value between 0 and 1.
\begin{table}[htb]
\centering
    \caption{Network structure of discriminator}
    \label{table:Discriminator}
    \begin{threeparttable}
    \begin{tabular}{|c|c|@{}c@{}|}
    \hline 
    Layers& Details & Output Size\\
    \hline
    Input Image & - & $64\times64\times1$\\
    \hline
    Padding &Padding3$\times$3 & $70\times70\times1$\\
    \hline
    Conv
        & 
        \begin{tabular}{c}
            Conv4$\times$4,stride=2\\ 
            LeakyReLU
        \end{tabular} 
        & 
        \begin{tabular}{c}
            $32\times32\times64$\\ 
            $32\times32\times64$
        \end{tabular}\\
    \hline
    Conv
        & 
        \begin{tabular}{c}
            Conv4$\times$4,stride=2\\ 
            Batch Normalization\\ 
            LeakyReLU
        \end{tabular} 
        & 
        \begin{tabular}{c}
            $16\times16\times64$\\ 
            $16\times16\times64$\\ 
            $16\times16\times64$\\ 
        \end{tabular}\\
    \hline
    Conv
        & 
        \begin{tabular}{c}
            Conv4$\times$4,stride=2\\ 
            Batch Normalization\\ 
            LeakyReLU
        \end{tabular} 
        & 
        \begin{tabular}{c}
            $8\times8\times128$\\ 
            $8\times8\times128$\\ 
            $8\times8\times128$\\ 
        \end{tabular}\\
    \hline
    Conv
        & 
        \begin{tabular}{c}
            Conv4$\times$4,stride=1\\ 
            Batch Normalization\\ 
            LeakyReLU
        \end{tabular} 
        & 
        \begin{tabular}{c}
            $8\times8\times128$\\ 
            $8\times8\times128$\\ 
            $8\times8\times128$\\ 
        \end{tabular}\\
    \hline
    Conv
        & 
        \begin{tabular}{c}
            Conv4$\times$4,stride=1\\ 
        \end{tabular} 
        & 
        \begin{tabular}{c}
            $8\times8\times1$\\ 
        \end{tabular}\\
    \hline
    Flatten
        & 
        -
        & 
        \begin{tabular}{c}
            $64$\\ 
        \end{tabular}\\
    \hline
    Dense
        & 
        -
        & 
        \begin{tabular}{c}
            $1024$\\ 
        \end{tabular}\\
    \hline
    Dense
        & 
        -
        & 
        \begin{tabular}{c}
            $1$\\ 
        \end{tabular}\\
    \hline
    
    \hline 
    \end{tabular}
    \end{threeparttable}
    \end{table}

\subsection{Generator}
Generator is used to convert malware feature distributions to their `real' malware versions. It takes the corresponding batch size malware feature distribution $\mathcal{M}$ as input. $\mathcal{M}$ is an $M$-dimensional normalized third-order texture image. The input tensor is fed into a multi-layer sequential model with weights $\theta_{g}$. In addition to containing multiple convolutional blocks, it also contains nine levels of residual block (ResBlock) \cite{kupyn2018deblurgan:}. Each ResBlock consists of a convolution layer, instance normalization layer, and ReLU activation \cite{Nair2010RectifiedLU}. The output of this network is denoted as $\mathcal{T}$, we define a smooth function $G$ is defined to receive gradient information from the substitute detector, as shown in Equation (\ref{element-wise_max})
\begin{equation}\label{element-wise_max}
	G_{\theta_{g}}(\mathcal{M})= mean(\mathcal{M}, \mathcal{T})
\end{equation}
where the $mean(\cdot,\cdot)$ to represent element-wise average operation.

Pay attention to a few details in Tab. \ref{table:Generator}. First, we use the padding operation. The most important thing is to keep the boundary information. If no padding operation is added, the edge nodes of the input image will only be convolved once, and the pixels in the center of the image will be rolled. If you accumulate multiple times, then the boundary pixels will lose a lot of meaning. Second, the addition of the residual network is used to solve the degradation problem caused by the deep network, which leads to the problem of the accuracy of the training data. Third, the generator adds a transposition convolution (also called deconvolution) to achieve image upsampling. The significance is to restore the small-scale high-dimensional feature map back to make a pixel-wise prediction and obtain information about each point.
\begin{table}[htb]
\centering
\caption{Network Structure of Generator}
\label{table:Generator}
\begin{threeparttable}
\begin{tabular}{|c|c|@{}c@{}|}
\hline 
Layers& Details & Output Size\\
\hline
Input Image & - & $64\times64\times1$\\
\hline
Padding &Padding3$\times$3 & $70\times70\times1$\\
\hline
Conv
	& 
	\begin{tabular}{c}
		Conv7$\times$7\\ 
		LeakyReLU
	\end{tabular} 
	& 
	\begin{tabular}{c}
		$64\times64\times64$\\ 
		$64\times64\times64$
	\end{tabular}\\
\hline
Conv
	& 
	\begin{tabular}{c}
		Conv3$\times$3,stride=2\\ 
		Batch Normalization
	\end{tabular} 
	& 
	\begin{tabular}{c}
		$32\times32\times128$\\ 
		$32\times32\times128$
	\end{tabular}\\
\hline
Conv
	& 
	\begin{tabular}{c}
		Conv3$\times$3,stride=2\\ 
		Batch Normalization
	\end{tabular} 
	& 
	\begin{tabular}{c}
		$16\times16\times256$\\ 
		$16\times16\times256$
	\end{tabular}\\
\hline
Padding & Padding3$\times$3 & $18\times18\times256$\\
\hline 
ResNet\tnote{1}
	& 
	\begin{tabular}{c@{}c}
		9$\times$
		&
		\begin{tabular}{l}
			Conv\tnote{2}\\ 
			Padding\\
			Conv\\ 
			Add\tnote{3}\\
			Padding\\
		\end{tabular}\\
	\end{tabular} 
	& 
	\begin{tabular}{c}
		$16\times16\times256$\\ 
		$18\times18\times256$\\
		$16\times16\times256$\\ 
		$16\times16\times256$\\
		$18\times18\times256$\\ 
	\end{tabular}\\
\hline
Up sampling & - & $32\times32\times256$\\
\hline
Conv
	& 
	\begin{tabular}{c}
		Conv3$\times$3\\ 
		Batch Normalization
	\end{tabular} 
	& 
	\begin{tabular}{c}
		$32\times32\times128$\\ 
		$32\times32\times128$
	\end{tabular}\\
\hline
Up sampling & & $64\times64\times128$\\
\hline
Conv
	& 
	\begin{tabular}{c}
		Conv3$\times$3\\ 
		Batch Normalization
	\end{tabular} 
	& 
	\begin{tabular}{c}
		$64\times64\times64$\\ 
		$64\times64\times64$
	\end{tabular}\\
\hline
Padding & Padding3$\times$3 & $70\times70\times64$\\
\hline
Conv & Conv7$\times$7 & $64\times64\times1$\\
\hline
Add 
	&
	\begin{tabular}{c} Input Image\\ Conv \end{tabular}
	& 
	$64\times64\times1$\\
\hline
Lambda & - & $64\times64\times1$\\

\hline 
\end{tabular}
\begin{tablenotes}
        \footnotesize
        \item[1] ResNet contains nine residual blocks, each of which contains two convolution operations, two padding operations, and one add operation. 
        \item[2] The Conv here contains a Conv layer with 3$\times$3 convolution kernel and the stride of 1 and a Batch Normalization operation. 
        \item[3] The Add layer connects information before and after residuals.
      \end{tablenotes}
\end{threeparttable}
\vspace{-8pt}
\end{table}
\section{Training MTFD-GANs}
The texture segmentation and tensor singular value decomposition data after malware mapping are enhanced and unified into a size of $64\times 64\times 1$. The label of the benign sample is defined as 0, the sample tag of the malware is 1, the pre-trained Black-Bone detector is used to predict the outcome of the benign and malware, and the score of the predicted classification result is correspondingly labeled. We set up the training epochs and a mini-batch. The training of the GANs is an automated game process. The generator is to generate more realistic malicious samples, which can fool the discriminator detection. The discriminator is used to better identify malware and benign software. In particular, the quality of the discriminator discriminating corresponds to the Black-Bone detector because the label input to the discriminator model training is the result of the Black-Bone detector prediction. Therefore, the process of optimizing the objective function parameters is to minimize the direction of the gradient and continuously update the parameter values of the generator and discriminator networks. Note that the parameter values of the discriminator network are fixed when training the generator network. 

The loss function of the discriminator in GANs is defined as $L_D$, see Equation (\ref{D_train}), which is the sum of the losses of malicious and benign software. 
\begin{equation}\label{D_train}
\begin{aligned}
		L_{D} &= L_{D}\_{benign} + L_{D}\_{mal} \\
        &= \mathbb{E}_{\boldsymbol{x} \thicksim BD_{benign}} [\log D_{\theta_D}(\mathcal{X})] + \mathbb{E}_{\boldsymbol{x} \thicksim BD_{mal}} \\
        &[\log(1-D_{\theta_D}(\mathcal{X}))] \\
        &= \sum_{\boldsymbol{x}}  BD_{benign}(\boldsymbol{x}) \log D_{\theta_D}(\mathcal{X}) + \sum_{\boldsymbol{x}}  B D_{mal}(\boldsymbol{x}) \\
        &\log(1-D_{\theta_D}(\mathcal{X}))
\end{aligned}
\end{equation}
where $BD(\cdot)$ denotes the Black-Bone detector, $\mathcal{X}$ is a tensor size of the input discriminator network of $64 \times 64 \times 1$, and $\theta_D$ represents a parameter of the discriminator network. The goal is to minimize discriminator $D_{min}$ while updating parameter $\theta_D$ in Equation (\ref{D_mintrain})
\begin{equation}\label{D_mintrain}
\begin{aligned} 
\nabla_{\theta_{D}}L{D} &=\nabla_{\theta_{D}}\mathbb{E}_{\boldsymbol{x}\thicksim BD_{benign}}[\log D_{\theta_D}(\mathcal{X})] \\ & +\nabla_{\theta_{D}} \mathbb{E}_{\boldsymbol{x} \thicksim BD_{mal}}
		[\log(1- D_{\theta_D}(\mathcal{X}))] \\
        &=\nabla_{\theta_{D}} \frac{1}{N} \sum_{\boldsymbol{x}}^{N}[BD_{\text {benign}}(\boldsymbol{x}) \log D_{\theta_D}(\mathcal{X})\\
        +&BD_{\text{mal}}(\boldsymbol{x})\log (1-D_{\theta_D}(\mathcal{X}))]
\end{aligned}
\end{equation}
We expect the discriminator to lose weight after the GANs training, which ensures that the black box detector can predict more accurate malware.

The generator mode loss function in GANs is defined as $L_G$, see Equation \ref{G_train}, which is a comparison to real malware tags.
\begin{equation}\label{G_train}
\begin{aligned} 
L_{G} &=  \mathbb{E}_{\boldsymbol{z} \thicksim mal_{z}} [log (1 - D_{\theta_D}(G_{\theta_G}(\mathcal{Z})))]\\
        &= \sum_{x} y_{true_{mal_{z}}}log (1 - D_{\theta_D}(G_{\theta_G}(\mathcal{Z})))
\end{aligned}
\end{equation}
where $y_{true_{mal_z}}$ is the label of real malware. The tensor $\mathcal{Z}$ is the `noise' that matches the distribution of malware samples. The size of the input to the generator network model is $64\times 64\times 1$. The goal is to maximize the generator $G_{max}$, fix the parameter $\theta_D$, and update the parameter $\theta_G$ in Equation (\ref{G_maxtrain}).
\begin{equation}\label{G_maxtrain}
\begin{aligned} 
\nabla_{\theta_{G}}L_{G} &= \nabla_{\theta_{G}} \mathbb{E}_{\boldsymbol{z} \thicksim mal_{z}} [log (1 - D_{\theta_D}(G_{\theta_G}(\mathcal{Z})))]\\
        &= \frac{1}{N} \sum_{x}^{N} y_{true_{mal_{z}}}log (1 - D_{\theta_D}(G_{\theta_G}(\mathcal{Z})))
\end{aligned}
\end{equation}
We want the generator to generate real malicious samples while deceiving the discriminator's detection and misleading it into benign software. The parameter optimization of the discriminator and generator model uses the gradient descent algorithm, which is commonly used to solve the unconstrained optimization problem, that is, to find the best function matching of the data by minimizing the square of the error function. 

We define perceptual loss $\ell_{\text {feat}}^{TrainedNet, j}$ in Equation (\ref{perceptual_loss}) to calculate the potential advanced feature difference between the original malware map and the newly generated malicious map of the GANs. $TrainedNet$ is a pre-trained network model that uses VGG-16 \cite{simonyan2015very}, which is a pre-training using the ImageNet dataset \footnote{https://www.sec.cs.tu-bs.de/$\thicksim$danarp/drebin/download.html}.
\begin{equation}\label{perceptual_loss}
\begin{aligned}
	&\ell_{\text {feat}}^{TrainedNet, j}(validity, mal) =\frac{1}{C_{j} H_{j} W_{j}} \\
	&\left\|TrainedNet_{j}(validity)-TrainedNet_{j}(mal)\right\|_{2}^{2}
\end{aligned}
\end{equation}
where $j$ represents the $j$-th layer of the network, $C_{j} H_{j} W_{j}$ represents the size of the feature map of the $j$-th layer, and validity represents the malicious map generated by the generator in GANs. The high-level feature loss function is used to preserve the global structure and content of the image, but the color texture and the exact shape no longer exist.

Algorithm \ref{algorithm:Train-MTFD-GANs} is the training process of our proposed MTFD-GANs model.
\begin{algorithm}[htb]
\caption{The Training Process of MTFD-GANs}\label{algorithm:Train-MTFD-GANs}
\begin{algorithmic}[1]
\REQUIRE Malware and tags ($X_{mal}$, $Y_{mal}$), benign and tags ($X_{benign}$, $Y_{benign}$)
\FOR{i $\gets$ 1 to epochs}  
    \STATE Sample a minibatch of $m$ malware samples from malware distribution $P_{mal}(\bf{z})$ \label{minibatch_mal}
	\STATE Sample a minibatch of $m$ benign softwares from benign distribution $P_{benign}(\bf{x})$ \label{minibatch_benign}
	\STATE Label malware and benign softwares with a trained Black-Bone detector and predict the accuracy of malware $ACC_0$ \label{label_black} 
	\STATE Update the discriminator $\theta_{d}$ by descending along the gradient: $\nabla_{\theta_{d}} L_{D}$ \label{update_D}
	\STATE Update the generator $\theta_{g}$ by descending along the gradient: $\nabla_{\theta_{g}} L_{G}$ \label{update_G}
\ENDFOR
\STATE Retrain the Black-Bone detector with newly generated malware in GANs \label{Retrain_black}
\STATE Detect malware accuracy $ACC_1$ using Black-Bone detector
\STATE Calculate the accuracy of the model lift: $Impr_{Ratio}=\frac{\|ACC_0-ACC_1\|}{\|ACC_0\|}$ \label{lift_acc}
\ENSURE Model (model.h5), parameters (weight.h5) and Improved accuracy ($IPA_{Ratio}$)
\end{algorithmic} 
\end{algorithm}

In line \ref{minibatch_mal} and line \ref{minibatch_benign}, the selected mini-batch is used for training sampling of malware and benign software. Then, in line \ref{label_black}, we use the Black-Bone detector to predict the classification result, and get the label value of the corresponding software (benign or malicious) and input it into the discriminator in GANs. Instead of using the original artificially tagged malware and benign software tag values, the generator does not have a training direction at first, and the `real' malware generated may not match the original malware characteristics distributed. And the accuracy of the predicted malware is used to compare the proposed MTFD-GANs model to improve the efficiency of malware detection after training, in lines \ref{Retrain_black} to \ref{lift_acc}. In line \ref{update_D} and line \ref{update_G}, we use the gradient descent algorithm to iteratively update the network layer parameters of the discriminator and generator in GANs.

\section{Experiments}

\subsection{Experiment Setup}
We comprehensively evaluate our method on the DREBIN, which is a dataset widely used for malware detection \cite{arp2014drebin}. This dataset contains 5560 files from 179 different malware families, which performs a broad static analysis, gathering as many features of an application as possible. We extract some datasets for evaluation experiments. After screening statistics, the number of malicious Android software selections is close to that of normal software selection, avoiding the problem of imbalance between positive and negative samples.

The .apk file for an Android application contains the main five parts: the .dex file directly executed by the Dalvik virtual machine, the manifest file Androidmanifest.xml placed in the configuration, and the third-party library file lib of the jar package added, the resource file res and META-INF file for storing signature information.
An Android application is commonly written in Java and compiled to Dalvik byte-code which contained in a .dex file. This file can be just-in-time compiled by the Dalvik virtual machine or compiled once into a system-dependent binary by ART on the Android platform. Our most important task is to decompress an Android .apk software to get the .dex file, which is a binary container for the byte-code and the data within the classes.

We use an Nvidia GeForce GTX 1080 GPU for development of the network. We adopt two ways to split the dataset. The first splitting way is to divide the original malware and benign dataset with their tags by $80\%$ and $20\%$. MTFD-GANs and the Black-Bone detector share the same training set. The second segmentation method selects $40\%$ of the data set as the training set for MTFD-GANs, selects another $40\%$ of the data set as the training set for the Black-Bone box detector, and the remaining $20\%$ of the data is used as the test set.
\subsection{Analyze malware generation efficiency}
We first analyze the ability of the new malware generated by the MTFD-GANs model. For better quantification, the Black-Bone detection accuracy (BBDA) indicates the accuracy of malware detection. The better result is that the malware generated after the MTFD-GANs training can successfully deceive the detection by the Black-Bone detector. In other words, the new malware generated by MTFD-GANs is different from the original malware data feature distribution.

We compare the accuracy of malware detection by the Black-Bone detector before and after MTFD-GANs training. On the one hand, the use of BBDA indicators reflects the efficiency of new malware generated by the generator of GANs. On the other hand, the use of tags predicted by malware detectors allows the generator's training to converge faster. Unlike the tags that use the original malware and benign software, the MTFD-GANs can't tend to do four kinds of classification learning at once (original and new generated malware as well as raw and new generated benign software). BBDA on the training set and the test set of original samples and training generated malware samples is shown in Table \ref{samedata_BBDA_table}. The accuracy of the original Black-Bone detector averaged over $90.4\%$. After learning the confrontational generation model, the malware detection rate was up to $23\%$. This shows that the new malware sample generated by the MTFD-GANs is different from the original malware feature distribution. This can effectively deceive the detection of the Black-Bone detector and reflect the efficiency of the generator model.

\begin{figure}[htb]\small
  \centering
  \includegraphics[width=0.48\textwidth]{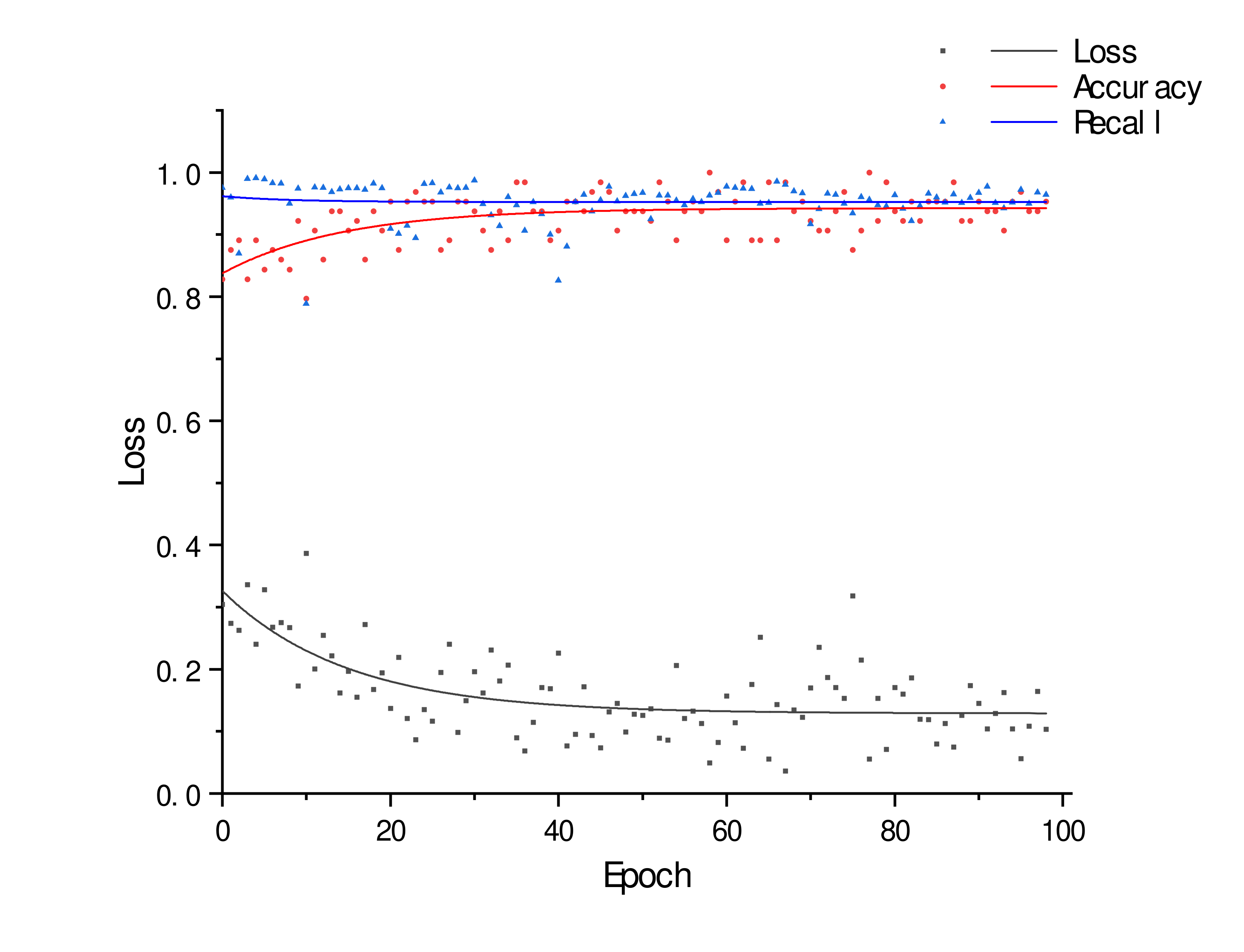}
  \caption{MTFD-GANs training based on the decision tree (DT) Black-Bone detection. The black points indicate the loss of MTFD-GANs training, the red points indicate the accuracy of the MTFD-GANs training, and the blue points indicate the recall rate of Black-Bone detection. The curves corresponding to the colors are their asymptotic fitting curves. }\label{fig:training image}
\end{figure}

\begin{table}[htb]\centering
\setlength{\abovedisplayskip}{4pt}
\caption{Malware detection accuracy (in percentage) on original samples and newly generated samples when MTFD-GANs and the Black-Bone detector are trained on the same training set. `Original' represents the result of malware detected by the Black-Bone detector before MTFD-GANs is not trained. `Trained' represents the result of malware detection after MTFD-GANs training. The Attention detector contains 100 training epoches.}\label{samedata_BBDA_table}
\begin{tabular}{|@{}c@{}|@{}c@{}|@{}c@{}|@{}c@{}|@{}c@{}|}
\hline
\multirow{2}{*}{ Black-Bone } & \multicolumn{2}{c|}{Training data}   & \multicolumn{2}{c|}{Test Data} \\ \cline{2-5} 
                                              & { Original } & { Trained } & { Original } & { Trained } \\ \hline
LR                                   & 99.97               & 23.32              & 91.84               & 22.79              \\ \hline
SVM                                  & 82.50               & 0.00               & 81.11               & 0.00               \\ \hline
MLP                                  & 92.51               & 2.02               & 83.77               & 1.58               \\ \hline
DT                                   & 96.44               & 8.17               & 95.09               & 7.73               \\ \hline
RF                                   & 86.79               & 0.37               & 85.35               & 0.24               \\ \hline
AdaBoost                             & 89.70               & 0.04               & 86.68               & 0.00               \\ \hline
GBDT                                 & 94.90               & 1.16               & 87.85               & 0.99               \\ \hline
Attention                            & 96.37               & 0.00               & 86.99               & 0.00               \\ \hline
NB                                   & 79.13               & 0.00               & 78.28               & 0.00               \\ \hline
\end{tabular}\vspace{-6pt}
\end{table}
In order to better observe the effectiveness of the entire MTFD-GANs training process, we select the Black-Bone detector with the decision tree (DT) as an example. The training results are shown in Fig. \ref{fig:training image}. We can clearly see that the accuracy of MTFD-GANs training is maintained at 92.70\%, and the error degree is close to 0.159159, and in the detection of whether the DT detector determines whether it is a malware sample, it has a high recall rate, and the average value is 95.24\%.

In real-world situations, the selection of training data in the MTFD-GANs and Black-Bone detector should be unshared and interfere with each other. Table \ref{differentdata_BBDA_table} shows that the average accuracy of malware detection based on model training on different training sets is $84.4\%$, which is lower than the model detection rate based on the same training set. Because there is model training between GANs and Black-Bone detectors, training the model with the same dataset will enable the model to learn more rich malware data features. However, if it is a different training dataset, there is a certain degree of error in the malware characteristics learned by the Black-Bone detector and the malware characteristics generated by the GANs training, which leads to a decrease in the accuracy of the model.
\begin{table}[htb]\centering
\setlength{\abovedisplayskip}{4pt}
\caption{Malware detection accuracy (in percentage) on original samples and newly generated samples when MTFD-GANs and the Black-Bone detector are trained on different training sets. `Original' represents the result of malware detected by the Black-Bone detector before MTFD-GANs is not trained. `Trained' represents the result of malware detection after MTFD-GANs training.  The Attention detector contains 100 training epoches.}\label{differentdata_BBDA_table}
\begin{tabular}{|@{}c@{}|@{}c@{}|@{}c@{}|@{}c@{}|@{}c@{}|}
\hline
\multirow{2}{*}{ Black-Bone } & \multicolumn{2}{c|}{Training data}    & \multicolumn{2}{c|}{Test Data}        \\ \cline{2-5} 
                                              & { Original } & { Trained } & { Original } & { Trained } \\ \hline
LR                                   & 100                 & 31.86              & 86.10               & 28.86              \\ \hline
SVM                                  & 85.39               & 0.04               & 81.28               & 0.08               \\ \hline
MLP                                  & 77.45               & 22.50              & 72.12               & 21.21              \\ \hline
DT                                   & 73.37               & 14.76              & 71.96               & 15.22              \\ \hline
RF                                   & 90.84               & 1.12               & 87.43               & 0.58               \\ \hline
AdaBoost                             & 92.13               & 0.12               & 86.18               & 0.00               \\ \hline
GBDT                                 & 97.83               & 0.66               & 86.68               & 0.41               \\ \hline
Attention                            & 68.14               & 0.00               & 70.79               & 0.00               \\ \hline
NB                                   & 78.95               & 0.00               & 78.20               & 0.00               \\ \hline
\end{tabular}\vspace{-6pt}
\end{table}

\subsection{Improve malware detection efficiency}
Newly generated malware samples enrich the set of potential data features in malware. If the newly generated malware is repeatedly trained in the black box detector, the efficiency of the model detection and the robustness of the model can be improved.

Table \ref{Compare MTFD-GANs with Black-Bone detection BBDA} shows that MTFD-GANs has improved against existing malicious code detection models. The data entered into the trained MTFD-GANs and black box detectors are the original malware that has not been trained. Using the cross entropy loss function, the model prediction results show that the MTFD-GANs has the largest increase in malware detection rate relative to 100 rounds of iterative training (Attention Model), which is close to $41.6\%$. The SVM model has the smallest improvement, which is nearly $1.31\%$. The reason why the detection rate of the RF model is higher than that of the MTFD-GANs may be caused by the mechanism inside the random forest. The RF model cannot control the entire internal operation, and can only try between different parameters and random seeds. There may be many similarities. The decision tree masks the real results. The detection rate of the BN model and the MTFD-GANs is basically the same, because the BN network has a very fatal problem. For the categorical variable features in the test set, if not seen in the training set, the probability of directly calculating the zero prediction function will be invalid.
\begin{table}[htb]\centering
\setlength{\abovedisplayskip}{4pt}
\caption{Comparison of malware detection accuracy (in percentage) between traditional Black-Bone detector and MTFD-GANs. The Attention detector contains 100 training epoches.}\label{Compare MTFD-GANs with Black-Bone detection BBDA}
\begin{tabular}{|c|c|l|c|l|}
\hline
\multirow{2}{*}{Black-Bone detector} & \multicolumn{4}{c|}{Real untrained malware}                                         \\ \cline{2-5} 
                                              & \multicolumn{2}{c|}{Black-Bone} & \multicolumn{2}{c|}{MTFD-GANs} \\ \hline
LR                                   & \multicolumn{2}{c|}{91.10}                 & \multicolumn{2}{c|}{98.19}                \\ \hline
SVM                                  & \multicolumn{2}{c|}{76.78}                 & \multicolumn{2}{c|}{77.28}                \\ \hline
MLP                                  & \multicolumn{2}{c|}{91.16}                 & \multicolumn{2}{c|}{97.85}               \\ \hline
DT                                   & \multicolumn{2}{c|}{71.96}                 & \multicolumn{2}{c|}{74.83}                \\ \hline
RF                                   & \multicolumn{2}{c|}{87.02}                 & \multicolumn{2}{c|}{76.12}                \\ \hline
AdaBoost                             & \multicolumn{2}{c|}{84.69}                 & \multicolumn{2}{c|}{89.26}                \\ \hline
GBDT                                 & \multicolumn{2}{c|}{87.85}                 & \multicolumn{2}{c|}{89.35}                \\ \hline
Attention                            & \multicolumn{2}{c|}{68.78}                 & \multicolumn{2}{c|}{97.34}                \\ \hline
NB                                   & \multicolumn{2}{c|}{68.55}                 & \multicolumn{2}{c|}{65.64}                \\ \hline
\end{tabular}\vspace{-6pt}
\end{table}



\section{Conclusion}
In this paper, we propose a flexible Android malware detection model based on GANs with code tensor (MTFD-GANs) to improve the detection efficiency of malware. We use an end-to-end malicious code mapping lossless coding method, propose an image texture cutting by segment self-growing method, and establish a location-based hash-based image retrieval algorithm to select code segments with significant malicious features. The low-tubal-rank tensor singular value decomposition algorithm (tSVD) is used to solve the problem of inconsistent cut image size. The experiment compares a variety of existing black box detection models, and the proposed models are superior to them in malware detection. Through the training of GANs, the generated malware features not only enrich the diversity of malware samples, but also enhance the robustness of malware detection models.



%

%
\input{Yang.bbl}

\bibliographystyle{IEEEtran}
\nocite{*} 

\end{document}

%% file: Yang.bbl

%% file: Yang.bbl
\begin{thebibliography}{10}
\providecommand{\url}[1]{#1}
\csname url@samestyle\endcsname
\providecommand{\newblock}{\relax}
\providecommand{\bibinfo}[2]{#2}
\providecommand{\BIBentrySTDinterwordspacing}{\spaceskip=0pt\relax}
\providecommand{\BIBentryALTinterwordstretchfactor}{4}
\providecommand{\BIBentryALTinterwordspacing}{\spaceskip=\fontdimen2\font plus
\BIBentryALTinterwordstretchfactor\fontdimen3\font minus
  \fontdimen4\font\relax}
\providecommand{\BIBforeignlanguage}[2]{{%
\expandafter\ifx\csname l@#1\endcsname\relax
\typeout{** WARNING: IEEEtran.bst: No hyphenation pattern has been}%
\typeout{** loaded for the language `#1'. Using the pattern for}%
\typeout{** the default language instead.}%
\else
\language=\csname l@#1\endcsname
\fi
#2}}
\providecommand{\BIBdecl}{\relax}
\BIBdecl

\bibitem{GdataReport}
G.~Data, ``Cyber attacks on android devices on the rise,'' 2018,
  https://www.gdatasoftware.com/blog/2018/11/31255-cyber-attacks-on-android-devices-on-the-rise.

\bibitem{Christodorescu2005SemanticsawareMD}
M.~Christodorescu, S.~Jha, S.~A. Seshia, D.~X. Song, and R.~E. Bryant,
  ``Semantics-aware malware detection,'' \emph{2005 IEEE Symposium on Security
  and Privacy (S\&P'05)}, pp. 32--46, 2005.

\bibitem{Yin2007PanoramaCS}
H.~Yin, D.~X. Song, M.~Egele, C.~Kr{\"u}gel, and E.~Kirda, ``Panorama:
  capturing system-wide information flow for malware detection and analysis,''
  in \emph{ACM Conference on Computer and Communications Security}, 2007.

\bibitem{Ye2017ASO}
Y.~Ye, T.~Li, D.~A. Adjeroh, and S.~S. Iyengar, ``A survey on malware detection
  using data mining techniques,'' \emph{ACM Comput. Surv.}, vol.~50, pp.
  41:1--41:40, 2017.

\bibitem{Cui2019MaliciousCD}
Z.~Cui, L.~Du, P.~Wang, X.~Cai, and W.~Zhang, ``Malicious code detection based
  on cnns and multi-objective algorithm,'' \emph{J. Parallel Distrib. Comput.},
  vol. 129, pp. 50--58, 2019.

\bibitem{Goodfellow2014GenerativeAN}
I.~J. Goodfellow, J.~Pouget-Abadie, M.~Mirza, B.~Xu, D.~Warde-Farley, S.~Ozair,
  A.~C. Courville, and Y.~Bengio, ``Generative adversarial nets,'' in
  \emph{NIPS}, 2014.

\bibitem{Xiao2018GeneratingAE}
C.~Xiao, B.~Li, J.-Y. Zhu, W.~He, M.~Liu, and D.~X. Song, ``Generating
  adversarial examples with adversarial networks,'' in \emph{IJCAI}, 2018.

\bibitem{Hu2017GeneratingAM}
W.~Hu and Y.~Tan, ``Generating adversarial malware examples for black-box
  attacks based on gan,'' \emph{CoRR}, vol. abs/1702.05983, 2017.

\bibitem{Gionis1999SimilaritySI}
A.~Gionis, P.~Indyk, and R.~Motwani, ``Similarity search in high dimensions via
  hashing,'' in \emph{VLDB}, 1999.

\bibitem{Kilmer2013Third}
M.~E. Kilmer, K.~Braman, N.~Hao, and R.~C. Hoover, ``Third-order tensors as
  operators on matrices: a theoretical and computational framework with
  applications in imaging,'' \emph{SIAM Journal on Matrix Analysis
  Applications}, vol.~34, no.~1, pp. 148--172, 2013.

\bibitem{kilmer2011factorizationGoogle}
M.~E. Kilmer and C.~D. Martin, ``Factorization strategies for third-order
  tensors,'' \emph{Linear Algebra and its Applications}, vol. 435, no.~3, pp.
  641--658, 2011.

\bibitem{Cortes1995SupportvectorN}
C.~Cortes and V.~Vapnik, ``Support-vector networks,'' \emph{Machine Learning},
  vol.~20, pp. 273--297, 1995.

\bibitem{Pampel2000LogisticRA}
F.~C. Pampel, ``Logistic regression: A primer,'' 2000.

\bibitem{Safavian1991ASO}
S.~R. Safavian and D.~A. Landgrebe, ``A survey of decision tree classifier
  methodology,'' \emph{IEEE Trans. Systems, Man, and Cybernetics}, vol.~21, pp.
  660--674, 1991.

\bibitem{Breiman2001RandomF}
L.~Breiman, ``Random forests,'' \emph{Machine Learning}, vol.~45, pp. 5--32,
  2001.

\bibitem{Hastie2005TheEO}
T.~J. Hastie, R.~Tibshirani, and J.~H. Friedman, ``The elements of statistical
  learning: data mining, inference, and prediction, 2nd edition,'' in
  \emph{Springer series in statistics}, 2005.

\bibitem{Bahdanau2015NeuralMT}
D.~Bahdanau, K.~Cho, and Y.~Bengio, ``Neural machine translation by jointly
  learning to align and translate,'' \emph{CoRR}, vol. abs/1409.0473, 2015.

\bibitem{Freund1995ADG}
Y.~Freund and R.~E. Schapire, ``A decision-theoretic generalization of on-line
  learning and an application to boosting,'' in \emph{EuroCOLT}, 1995.

\bibitem{Friedman1991MultivariateAR}
J.~H. Friedman, ``Multivariate adaptive regression splines,'' \emph{Annals of
  Statistics}, vol.~19, no.~1, pp. 1--67, 1991.

\bibitem{Maron1961AutomaticIA}
M.~E. Maron, ``Automatic indexing: An experimental inquiry,'' \emph{J. ACM},
  vol.~8, pp. 404--417, 1961.

\bibitem{nataraj2011malware}
L.~Nataraj, S.~Karthikeyan, G.~Jacob, and B.~S. Manjunath, ``Malware images:
  Visualization and automatic classification,'' in \emph{International
  Symposium on Visualization for Cyber Security}, 2011.

\bibitem{abdulhayoglu2018use}
M.~A. Abdulhayoglu and B.~Thijs, ``Use of locality sensitive hashing (lsh)
  algorithm to match web of science and scopus.'' \emph{Scientometrics}, vol.
  116, no.~2, pp. 1229--1245, 2018.

\bibitem{Wang2003SupervisedTS}
B.~Wang and L.~Zhang, ``Supervised texture segmentation using wavelet
  transform,'' \emph{International Conference on Neural Networks and Signal
  Processing, 2003. Proceedings of the 2003}, vol.~2, pp. 1078--1082 Vol.2,
  2003.

\bibitem{Charalampidis2002WaveletbasedRI}
D.~Charalampidis and T.~Kasparis, ``Wavelet-based rotational invariant
  roughness features for texture classification and segmentation,'' \emph{IEEE
  transactions on image processing : a publication of the IEEE Signal
  Processing Society}, vol. 11 8, pp. 825--37, 2002.

\bibitem{Pichler1998AnUT}
O.~Pichler, A.~Teuner, and B.~J. Hosticka, ``An unsupervised texture
  segmentation algorithm with feature space reduction and knowledge feedback,''
  \emph{IEEE transactions on image processing : a publication of the IEEE
  Signal Processing Society}, vol. 7 1, pp. 53--61, 1998.

\bibitem{Haralick1973TexturalFF}
R.~M. Haralick, K.~S. Shanmugam, and I.~Dinstein, ``Textural features for image
  classification,'' \emph{IEEE Trans. Systems, Man, and Cybernetics}, vol.~3,
  pp. 610--621, 1973.

\bibitem{lecun1998gradient-based}
Y.~Lecun, L.~Bottou, Y.~Bengio, and P.~Haffner, ``Gradient-based learning
  applied to document recognition,'' \emph{Proceedings of the IEEE}, vol.~86,
  no.~11, pp. 2278--2324, 1998.

\bibitem{ioffe2015batch}
S.~Ioffe and C.~Szegedy, ``Batch normalization: Accelerating deep network
  training by reducing internal covariate shift,'' \emph{international
  conference on machine learning}, pp. 448--456, 2015.

\bibitem{xu2015empirical}
B.~Xu, N.~Wang, T.~Chen, and M.~Li, ``Empirical evaluation of rectified
  activations in convolutional network,'' \emph{arXiv: Learning}, 2015.

\bibitem{robinson2007explaining}
A.~Robinson, P.~S. Hammon, and V.~R. De~Sa, ``Explaining brightness illusions
  using spatial filtering and local response normalization,'' \emph{Vision
  Research}, vol.~47, no.~12, pp. 1631--1644, 2007.

\bibitem{kupyn2018deblurgan:}
O.~Kupyn, V.~Budzan, M.~Mykhailych, D.~Mishkin, and J.~Matas, ``Deblurgan:
  Blind motion deblurring using conditional adversarial networks,''
  \emph{computer vision and pattern recognition}, pp. 8183--8192, 2018.

\bibitem{Nair2010RectifiedLU}
V.~Nair and G.~E. Hinton, ``Rectified linear units improve restricted boltzmann
  machines,'' in \emph{ICML}, 2010.

\bibitem{simonyan2015very}
K.~Simonyan and A.~Zisserman, ``Very deep convolutional networks for
  large-scale image recognition,'' \emph{international conference on learning
  representations}, 2015.

\bibitem{arp2014drebin}
D.~Arp, M.~Spreitzenbarth, M.~Hubner, H.~Gascon, and K.~Rieck, ``Drebin:
  Effective and explainable detection of android malware in your pocket,'' in
  \emph{NDSS}, 2014.

\end{thebibliography}
